\documentclass[final,3p,times,twocolumn]{elsarticle}

\usepackage{graphics}
\usepackage{amssymb,amsmath}
\usepackage{color}

\usepackage{lineno}

\journal{Phys. Lett. B}

\begin{document}

\begin{frontmatter}

\title{Measurement of CNGS muon neutrino speed with Borexino}

\author[cern]{P. Alvarez Sanchez}

\author[polimi]{R. Barzaghi}

\author[mil]{G. Bellini}
\author[puchem]{J. Benziger}

\author[polimi]{B. Betti}
\author[polimi]{L. Biagi} 

\author[ham]{D.~Bick}
\author[lngs]{G.~Bonfini}
\author[vt]{D.~Bravo}
\author[mil]{M.~Buizza Avanzini}
% ...

\address[cern]{CERN, Geneva, Switzerland}
\address[polimi]{DIIAR-Politecnico di Milano, Piazza Leonardo da Vinci 32, 20133 Milano (ITALY)}
\address[mil]{Dipartimento di Fisica, Universit\'{a} degli Studi e INFN, Milano 20133, Italy}
\address[puchem]{Chemical Engineering Department, Princeton University, Princeton, NJ 08544, USA}
\address[ham]{University of Hamburg, Hamburg, Germany}
\address[lngs]{INFN Laboratori Nazionali del Gran Sasso, Assergi 67010, Italy}
\address[vt]{Physics Department, Virginia Polytechnic Institute and State University, Blacksburg, VA 24061, USA}
% ...

\begin{abstract}
We have measured the speed of muon neutrinos with the Borexino detector using short--bunch CNGS beams. The final result for the difference in time--of--flight between a $<$E$>$=17 GeV muon neutrino and a particle moving at the speed of light in vacuum is $\delta$t = 0.8 $\pm$ 0.7$_{stat}$ $\pm$ 2.9$_{sys}$ ns, well consistent with zero. 

\end{abstract}

\end{frontmatter}

\linenumbers

\section*{Introduction}
This paper describes a precise measurement of the speed of CNGS~\cite{bib:cngs} muon neutrinos made with the Borexino~\cite{bib:bx08det,bib:bxmuondet,bib:bxfilling} detector at the Laboratori Nazionali del Gran Sasso (LNGS) in Italy.

CNGS neutrinos travel about 730 km in matter with one of the highest relativistic $\gamma$ factors ever artificially produced. The neutrino mass is at most $\simeq$ 2 eV/c$^2$ or possibly much less, while the CNGS average beam energy is 17 GeV, so $\gamma$ is always $>$$10^{11}$, much bigger than that obtained in any charged particle beam. A test of Special Relativity with these particles is therefore meaningful. Besides, the measurement may also put an upper limit on the effect of non-standard propagation of neutrinos in matter.

This effort was also motivated by the claim, made by the Opera Collaboration in Sep. 2011 \cite{bib:opera}, that CNGS neutrinos travel faster than the speed of light in vacuum. This claim, however, was later withdrawn \cite{bib:opera12}.

If the mass of the heaviest neutrino is assumed to be 2 eV/c$^2$, the best direct limit on a neutrino mass, then the relativistic velocity of a 17 GeV neutrino should satisfy $|v-c|$/c $\lesssim$ 10$^{-19}$. Cosmological measurements~\cite{bib:cosmology} and neutrinoless double beta decay measurements~\cite{bib:dbd} give a mass limit an order of magnitude smaller, implying an even smaller constraint. However, theories with extra dimensions \cite{bib:theories0} predict apparent velocities different than the speed of light. Some of these theories \cite{bib:theories1, bib:theories2, bib:theories3} allow $|$v-c$|$/c $\simeq$ 10$^{-4}$ at neutrino energies of a few GeV. The present work has the sensitivity to test these theories and put limits on them. 

The MINOS collaboration has already performed a measurement with the NuMI beam, yielding a result compatible with c with an uncertainty of $\simeq$10$^{-4}$~\cite{bib:minos}. A similar measurement was also recently done by the Icarus collaboration \cite{bib:icarus} with a small sample of events collected in the Nov. 2011 CNGS run. 

We have been collecting CNGS events since May 2007, when the Borexino experiment started taking data. However, the triggering and time tagging systems originally designed for Borexino~\cite{bib:laben}, while perfectly adequate for solar neutrino physics~\cite{bib:solar1,bib:solar2,bib:solar3} and supernovae detection, were not sufficiently precise to allow an interesting measurement of the neutrino velocity. We have
therefore designed and installed a new facility capable of achieving an uncertainty of the order of a few ns. This new facility is based on a small--jitter analogue trigger, a geodetic GPS receiver, and a GPS-disciplined atomic Rb clock. The reader interested in the technical details and the calibration of this system may refer to~\cite{bib:hptf}.

The Borexino detector is a high--purity liquid scintillator calorimeter (within a Stainless Steel Sphere of $\approx$1300 m$^3$, SSS) shielded by a large Water Tank (WT, $\approx$3500 m$^3$) that serves also as a muon detector. It is installed in Hall C of the LNGS at a depth of 3800 m.w.e.

CNGS muon neutrinos are detected via charged current interactions that mostly occur in the rock upstream the detector. Internal events exist, but they cannot be easily disentangled from the crossing muons. Their number, however, is small ($\simeq$ 5\%). 

The kinematics of the muons does not affect the precision of the measurement. The difference in the time--of--flight of a $\approx$10 GeV muon with respect to a neutrino traveling a length of about 50 m (the approximate average distance traveled by such a muon in the rock plus the distance of the Borexino detector from the North wall of the Hall C) is less than 0.1 ns and can therefore be neglected. The same argument applies to the pions generated in the CNGS target at CERN. 

Muons can be detected by Borexino using the Water Tank (Cherenkov light) or the scintillator detector. In this analysis we have used only events detected by the latter since its precision is higher. Although only the core 270 t of PC-PPO scintillator are normally used for solar and geo-neutrino physics~\cite{bib:geo}, the sensitive mass for CNGS muon neutrinos is made by the whole $\approx$1300 m$^3$ of liquid scintillator and buffer liquid because, though quenched by a small amount of DMP (3 g/L), the buffer liquid light yield is sufficient to detect a muon ($\simeq$2 MeV/cm, equivalent in the buffer to approximately $\simeq$50~p.e./cm) .

The cross-section of the WT is 266 m$^2$ while that of the SSS is 147 m$^2$. These large areas yield a large number of events: in May 2012 we have collected a total of 291 events, 144 crossing the SSS and 147 crossing the WT only.

The distance of the SSS from the CNGS proton target was determined by performing a complete geodetic determination of the position in the IGS08 Reference Frame of the underground laboratory and of the Borexino detector in particular. The measurement was done in collaboration with LVD and ICARUS. The position of the CERN target was not measured again.  We rely on the precise knowledge of the location of each element of the accelerator already available at CERN. 

\begin{figure}[t]
\begin{center}
\includegraphics[width = 1.1\linewidth]{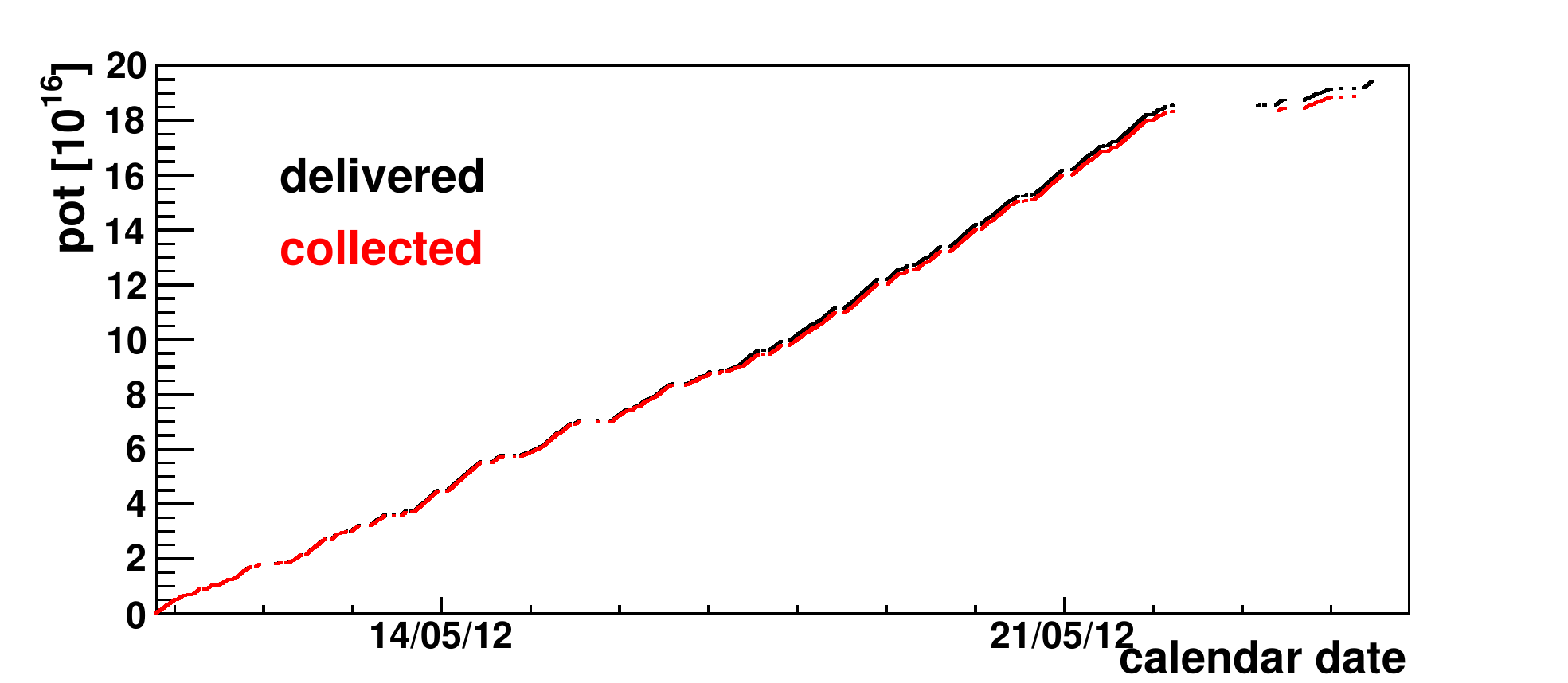}
\includegraphics[width = 1.1\linewidth]{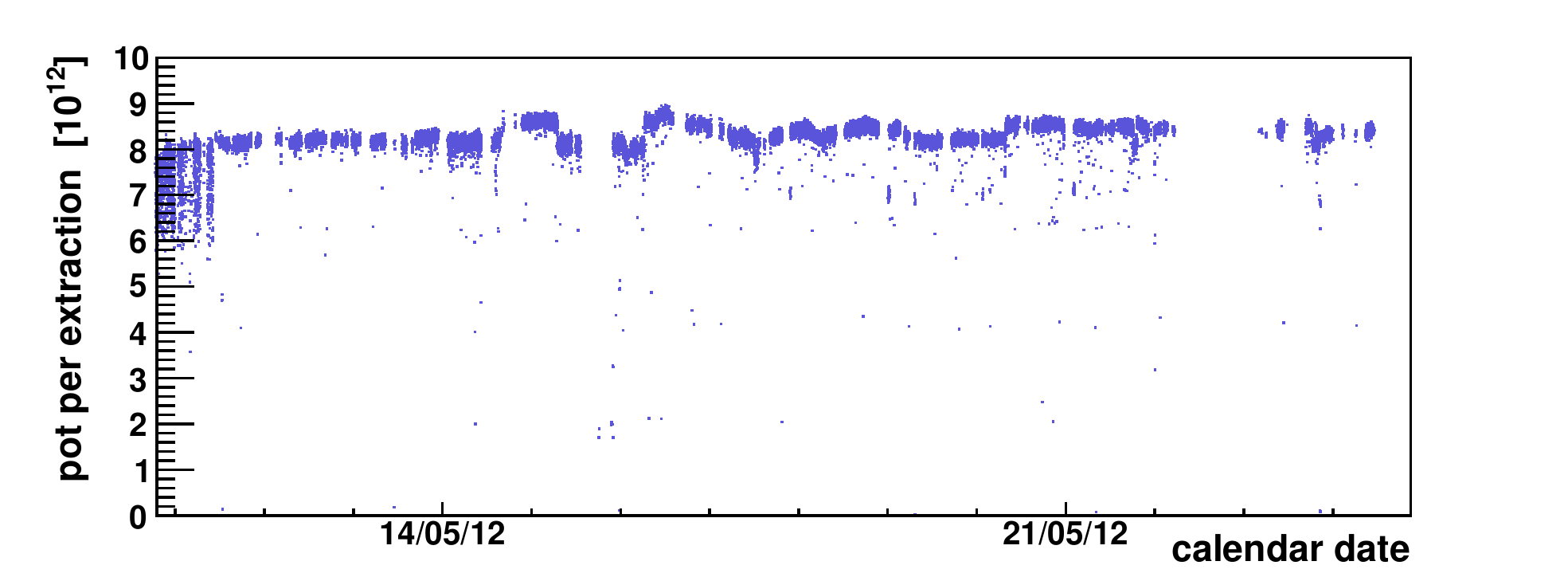}
\caption{Top: Cumulative distribution of pots delivered by CNGS and the equivalent number of pots collected during May 2012 run. The corresponding live--time is 97\%. Bottom: The pot delivered as a function of time during May 2012 run. }
\label{fig:pots}
\end{center}
\end{figure}

The paper is structured as follows: in next section we describe the measurement of the distance flown by the neutrinos; we then describe the analysis of the data collected in May 2012; for completeness, we also briefly report about the result obtained (with much less precision) in Oct.-Nov 2011. Finally, we report our result. 

\section*{Geodesy}
\label{sec:geodesy}
The distance between the Borexino reference point at LNGS and the target at CERN has been computed based on a geodetic survey at Gran Sasso Laboratories and the existing target co-ordinates at CERN. 

The geodetic campaign at LNGS was carried out in May 2012 and was performed in two steps. In the first step two GPS networks, one local and one regional, were established. 
The regional network used in this work consists of 32 GPS permanent stations located between CERN and Gran Sasso (including two antennas at CERN and one at LNGS) whose position in the IGS08 reference frame has been precisely estimated by adjusting (fitting) two weeks of GPS data (Bernese software \cite{bib:dach}, GPS weeks 1684 and 1685). 

A local 9 points GPS network has been monumented (installed) in the Gran Sasso area and framed to the regional one through 3 common stations using a four days-24 hours campaign. 
In the second step, based on the previously estimated GPS points, a high precision traverse has been measured along the Gran Sasso tunnel highway (10.5 km). This allowed the co-ordinates estimation of the Borexino reference point inside the LNGS. 

The IGS08 co-ordinates of this point and the related standard deviations are: (X= 4582157.919 m, Y= 1106469.341 m, Z=4283636.931 m); ($\sigma_X$=0.020 m, $\sigma_Y$=0.021 m, $\sigma_Z$=0.015 m). 
The co-ordinates and the precision of the target point at CERN, coded as TT41\_T\_40S, have been supplied by the CERN geodetic team. After datum shift to IGS08, assuming target point co-ordinates precision of 0.030 m, the geometric distance between the Borexino reference point at LNGS and the target at CERN has been estimated in 730472.082 $\pm$ 0.038 m. The error should be considered as 1 $\sigma$. 

\section*{Analysis of May 2012 bunched beam data}
\label{sec:may2012}
A special beam was set up at CERN in May 2012, optimized for the neutrino speed measurement. 

The main parameters of this beam are: narrow bunches ($\sigma \approx $2 ns), 16 bunches per batch with a bunch separation of $\simeq$100 ns, 4 batches per extraction separated by $\simeq$300 ns; the bunch intensity is $\simeq$10$^{11}$ protons; one extraction per CNGS cycle, with a cycle length of 13.2 s.

We have taken data from May 10$^{th}$ until May 24$^{th}$, 2012. Fig. \ref{fig:pots} top shows the cumulative distribution of the delivered protons on target (pots) and the equivalent pots collected by Borexino. The total pots delivered was 19.44 10$^{16}$, out of which 18.88 10$^{16}$ were recorded. The data taking efficiency during this period was 97\%. The beam delivery was very stable (see Fig. \ref{fig:pots} bottom). The missing data on May 23$^{rd}$ is due to SPS maintenance. 

As already mentioned in the introduction, we have collected a total of 291 events of which 144 crossed the SSS. The events were selected by requiring them to be on--time within a window of 100 $\mu$s with respect to the closest CNGS GPS tag plus 2.439 ms, the nominal neutrino time-of-flight. 

This large statistics allows us to apply stringent quality cuts and select the best data sample for the measurement. We have chosen not to use the WT only events, because the large size of the tank and the very complex internal geometry makes the full simulation of the light propagation and collection hard. The event sample was further reduced by the requirements of the High Precision Timing Facility (HPTF, \cite{bib:hptf}), which is a purely analogue trigger with a threshold of 100 mV, which corresponds to $\approx$800 p.e., i.e. $\approx$8 cm of muon track. This threshold reduces the acceptance of muons when they cross a small segment of the SSS. 

Borexino has developed muon reconstruction software capable of determining the location of the entrance point of a muon in the SSS with a precision of about 50 cm. This makes it possible to correct for the spherical shape of the detector which of course implies a different time--of--flight for different entrance points. 

This correction reduces the size of the data sample because some events cannot be properly reconstructed, but narrows the time distribution significantly, improving the quality of the measurement. After this final reduction, our data set consists of 
62 CNGS events, more than enough for a precision measurement.

The HPTF provides the GPS time of each trigger. The time-link between the CERN tagging system and the Borexino HPTF at Gran Sasso has been computed and calibrated by collaborators of the Italian and Spanish institutes of Metrology (INRIM and ROA, respectively) using Precise Point Positioning (PPP)~\cite{bib:PPP} and P3~\cite{bib:P3} algorithms, with latter implementing an All-In-View approach~\cite{bib:AV}. The calibration of the time link has been achieved with an uncertainty of 1.1 ns, being this the sum of a systematic uncertainty inherent to GPS system and available algorithms with an additional statistical uncertainty of 0.3 ns and 1 ns for the PPP and P3 algorithms, respectively. The procedure adopted to  calibrate the HPTF itself and the timeÐlink to CERN is fully described in~\cite{bib:hptf}. 

\begin{figure}[t]
\begin{center}
\includegraphics[width = \linewidth]{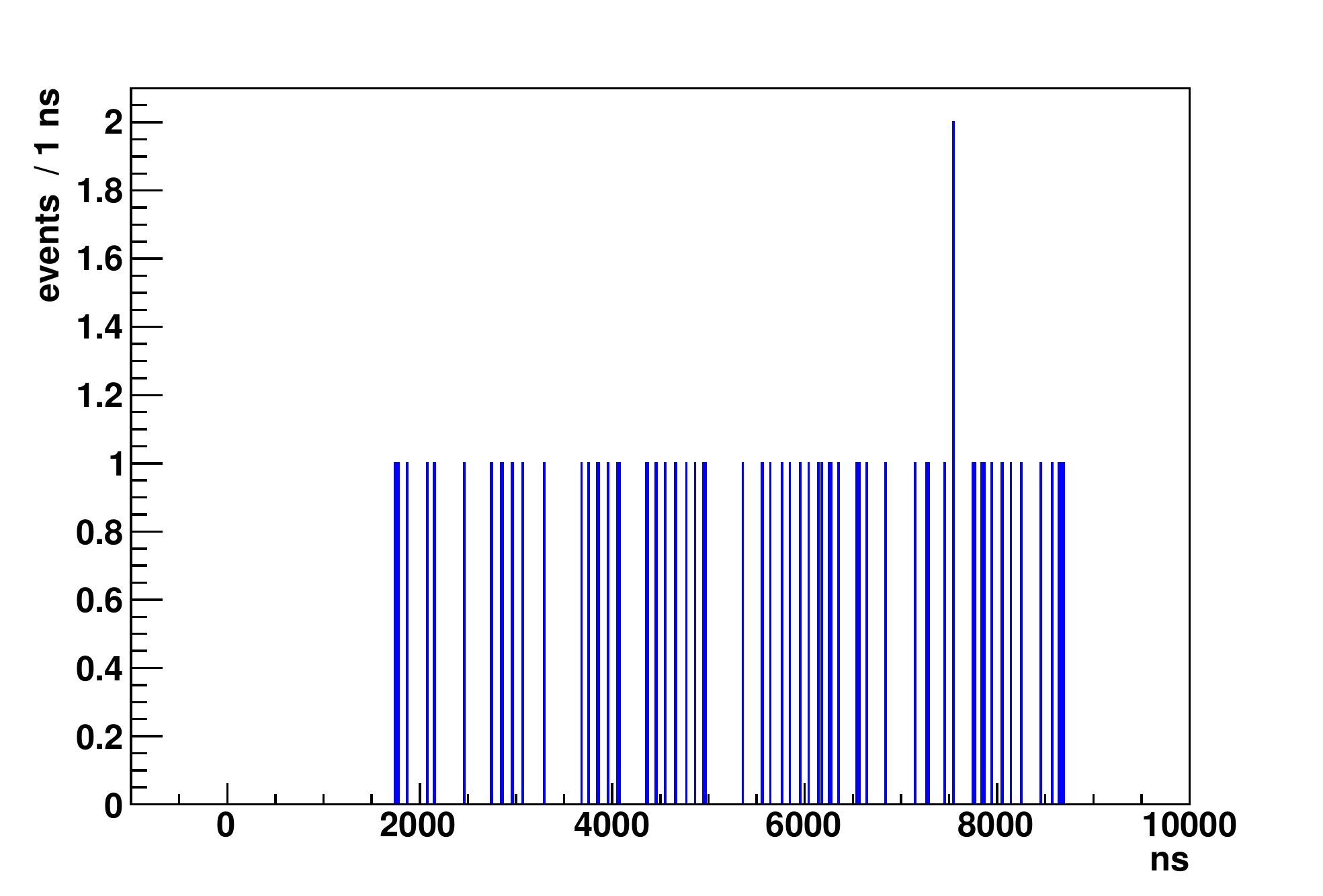}
\caption{Time distribution of the 98 inner detector events before folding of the 64 peaks and before the muon cuts. The horizontal axis is the time distance with respect to the first sample of the waveform digitizer described in the text. The distribution is flat and its total width is consistent with the expected 6900 ns. }
\label{fig:unfolded}
\end{center}
\end{figure}
Fig. \ref{fig:unfolded} shows the time distribution of the collected events before the folding of the peaks described below. The 98 events in the plot are those that survive a charge cut ($>$800 p.e., corresponding to approximately 2 MeV equivalent energy deposited in the scintillator) and that have valid data in the CNGS database. 

The time structure and the intensity of the proton bunches are recorded shot by shot through a Beam Current Transformer 
(BCT)\footnote{TT40:BCTFI.400344} located before the CNGS neutrino target ($743.391 \pm 0.002$ m upstream).

The signal is acquired by a fast (1 Gs/s) digitizer and stored in a database. We call  this "proton waveform."
Every proton waveform is time-tagged with respect to the CERN SPS timing system. 
The GPS time stamp marks the time of the first sample of the digitized signal. 
A detailed description of the CERN timing system and its calibration is available through public CERN reports \cite{bib:wiki}.

The time stamp associated with each waveform, after correction by some delays, defines the start time of the velocity measurement. Some of these delays are fixed in time while some of them are time dependent. 

The arrival time in Borexino is provided by the HPTF and the time-of-flight is obtained by the difference of the
two time tags plus a set of corrections due to: CERN-HPTF time link computed with PPP/P3 algorithms, fixed delays at CERN and Gran Sasso, and light propagation and trigger formation in Borexino (event dependent).

\begin{figure}[t]
\begin{center}
\includegraphics[width = 1.1\linewidth]{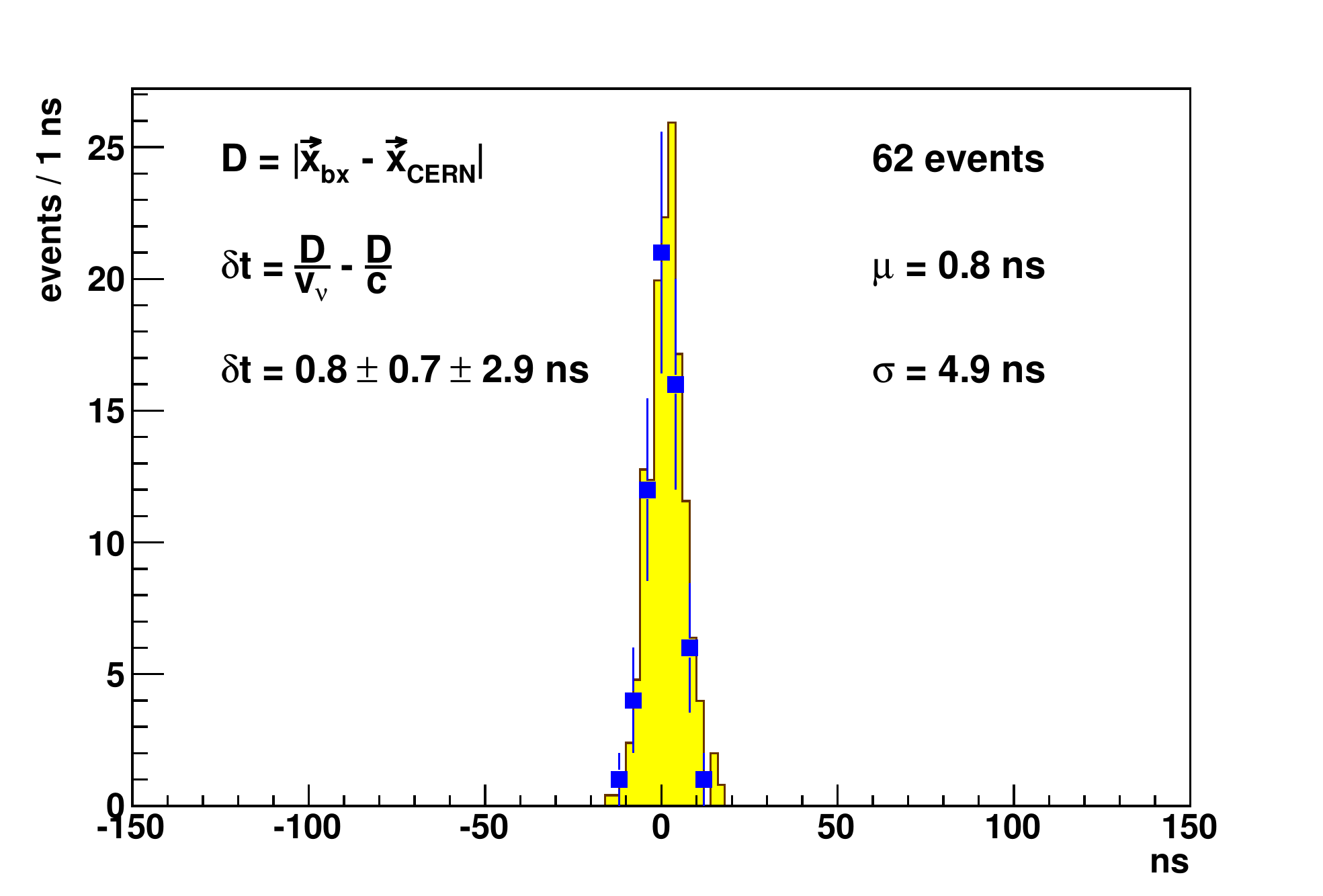}
\caption{Final distribution of the difference between the neutrino time--of--flight and that expected for a particle moving at speed c (data points). The mean value is consistent with zero and the width agrees with Monte Carlo simulation of known time jitters (yellow filled histogram). 
The small asymmetry is predicted by Monte Carlo and is due to the spherical shape of the Borexino
detector. The asymmetry does not affect the estimation of the central value by more than 200 ps. 
The mean value and its statistical error are obtained from the gaussian fit. The systematic error is
explained in the text. D is the distance between the proton target at CERN and the center of the Borexino detector. }
\label{fig:folded}
\end{center}
\end{figure}

The event distribution in Fig. \ref{fig:unfolded} is folded into a single one by using the information provided by the BCT digitized waveforms. After the corrections due to known instrumental delays, the geometrical correction that depends on the muon entrance point, the subtraction of the expected time--of--flight at speed c, and a 2.2 ns (positive) correction due to the Sagnac effect we obtain the distribution shown in Fig. \ref{fig:folded}. The distribution of the final sample of 62 events is well centered around zero and has a width of 4.9 ns. 

The width of the time-of-flight distribution is well understood. The light propagation in the detector, the collection of the light by the PMTs, and the formation of the trigger occur at a time that depends on the impact parameter of the muon track. Although we correct for the muon entrance point, the resolution of the reconstruction and the fluctuations in the propagation and in the collection of the light yield a distribution of finite width. We have therefore carefully simulated the broadening of the distribution due to the known effects with both the Geant-4 simulation and with a dedicated fast Monte Carlo. The result of both simulations is in good agreement with the data. The width of the arrival time distribution of the events from Monte Carlo, including light propagation, trigger formation, the jitter of some electronics modules, and the intrinsic 2 ns width of the bunches is shown in Fig. \ref{fig:folded}, overlaid with data. This Monte Carlo predicts a width of 5.1 ns, in very good agreement with the observed width of 4.9 ns. 

\section*{Analysis of Oct.-Nov. 2011 bunched beam data}
\label{sec:oct2011}

In the period from  Oct. 21$^{st}$ until Nov. 6$^{th}$, 2011 the standard CNGS beam was modified to produce 4 short bunches ($\sigma\approx$2 ns) separated by 524 ns. The intensity was 2.5 10$^{11}$ protons per bunch. 
In 16 days of operations CERN delivered 7.43 10$^{16}$ pots which we collected with an overall efficiency of 93\% (6.87 10$^{16}$ pots).

We took data with the standard triggering and GPS tagging systems, so the precision of this measurement is much lower than the one obtained in May 2012. In particular, the standard Borexino trigger has a relatively large intrinsic jitter of 32 ns due to the combined effect of a set of FPGAs and a DSP which limit the final precision. Besides, the absolute calibration of the GPS receiver delay in Hall C (ESAT-100 slave) has an uncertainty of a few ns.

The initial sample was made of 116 on--time events selected using both detectors in a time window of 16 $\mu$s around the nominal time--of--flight value. After selection of internal detector events, pile-up removal and quality cuts a sample of 36 events was used to measure the speed of neutrinos.

\begin{figure}[t]
\begin{center}
\includegraphics[width = 1.1\linewidth]{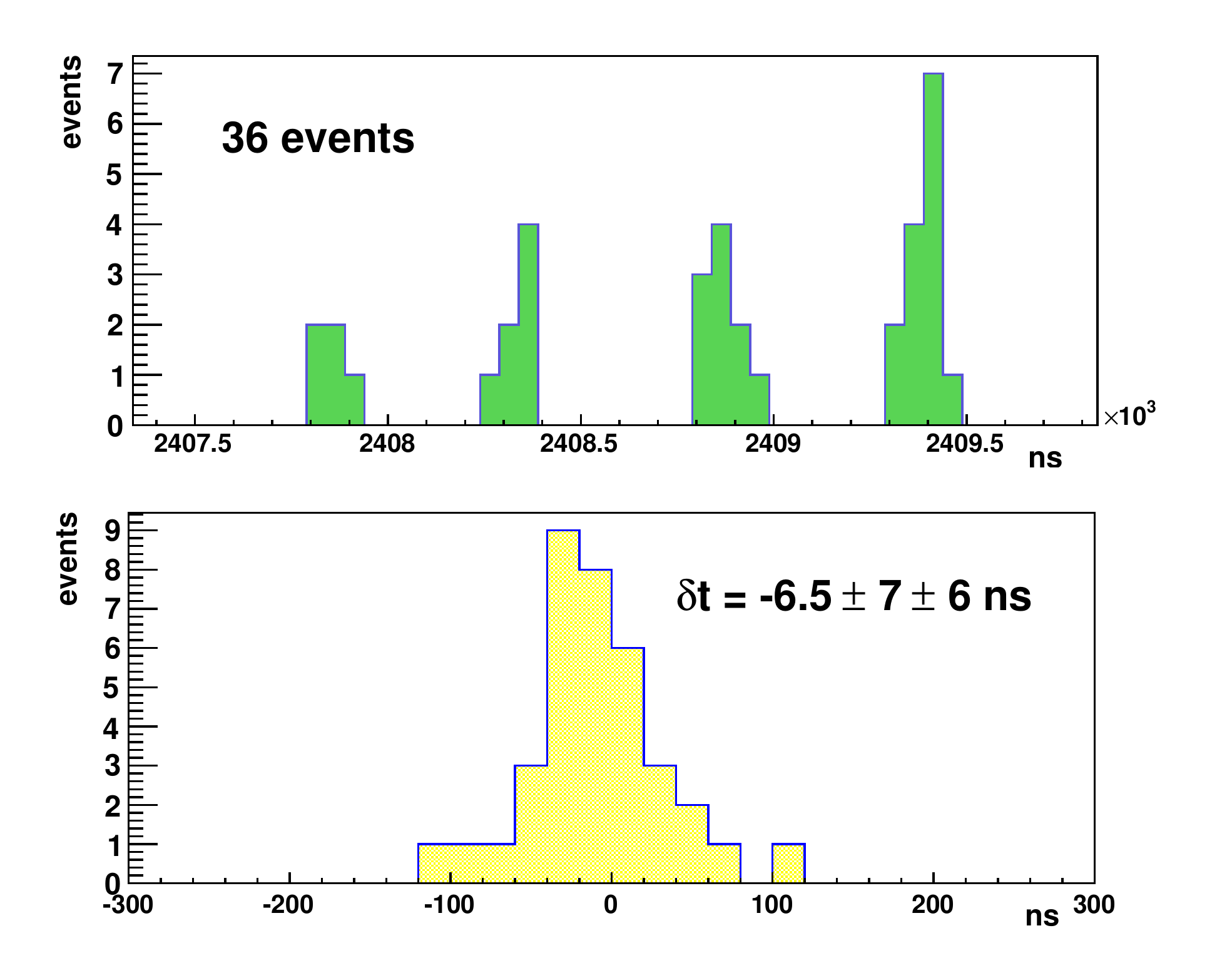}
\caption{Top: the time distribution of the CNGS neutrino candidates selected with the Inner Detector only from Oct.-Nov. 2011 bunched beam data. The four peaks separated by 524 ns are clearly visible. Bottom: the time--of--flight difference with respect to the speed of light after all corrections have been applied, including the Sagnac effect. The result is consistent with zero. }
\label{fig:old2011}
\end{center}
\end{figure}

We have applied to the final sample an analysis procedure very similar to that performed for the May 2012. The result is shown in Fig. \ref{fig:old2011}. As it is clear from the figure, the central value is well compatible with zero, but the error is much larger than the one obtained with May 2012 data. The statistical error comes from the width of the distribution. The systematic error, besides the contributions described in the next section, is bigger because of the poor knowledge of the ESAT-100 GPS receiver calibration and internal delays. 

\section*{Results}
\label{sec:results}
The final result obtained in May 2012 for the time--of--flight difference of $<$E$>$=17 GeV muon neutrinos with respect to the speed of light is $\delta$t = 0.8 $\pm$ 0.7$_{stat}$ $\pm$ 2.9$_{sys}$ ns, consistent with zero. The result obtained in Oct.-Nov. 2011 is also consistent with zero, but it is significantly less precise. Taking into account the fact that some of the errors affecting the two measurements are correlated (the CERN side is the same and some Borexino detector systematics are also the same) we have decided not to average the two numbers and keep May 2012 data only for the final result.

The statistical error comes from the fit to the distribution shown in Fig. \ref{fig:folded}. The systematic error depends on many sources that are listed and quantified in the Table below.

\begin{table}[h]
\begin{center}
\begin{tabular}{|l|c|} \hline\hline
$~~~~~~~~~~~~~~~~~~~$Description  &  Error (ns)  \\ \hline 
Time-Link Calibration (GPS)               &    1.1     \\
Borexino electronics delays                 &    0.5    \\
Delays at CERN                                   &    2.2     \\
Light propagation in BX detector         &    1.0     \\
Electronics resolution                          &     0.5     \\
Event selection stability                       &     1.0     \\
Geodesy measurement                      &      0.1     \\ \hline
Total systematic error                         &     2.9    \\  \hline \hline
\end{tabular}
\end{center}
\caption{Table of systematic errors contributing to the May 2012 measurement.}
\end{table}

\section*{Conclusions}
We have measured the speed of CNGS muon neutrinos with the Borexino detector. 

Setting D = $|\vec{x}_{bx} - \vec{x}_{CERN}|$=730472.082 $\pm$ 0.038 m as the distance between the target at CERN and the center of the Borexino detector, the final result for the relative time--of--flight distance between $<$E$>$=17 GeV muon neutrinos with respect to the speed of light,
$$\delta t = t_{BX} - t_{CERN} - \frac{D}{c}$$
is $\delta$t = 0.8 $\pm$ 0.7 $\pm$ 2.9 ns, consistent with zero. 
From this number we obtain $|$v-c$|$/c $<$ 2.1 10$^{-6}$ 90\% C.L..

\section*{Acknowledgements}
This work was funded by INFN (Italy), NSF (US Grant NSFPHY-0802646), BMBF (Germany), DFG (Germany, Grant OB160/1-1 and Cluster of Excellence "Origin and Structure of the Universe"), MPG (Germany), Rosnauka (Russia, RFBR Grant 09-02-92430), and MNiSW (Poland).  

We gratefully acknowledge the generous support of the Laboratori Nazionali del Gran Sasso, and we thank G. Di Carlo for valuable help in the calibration of the standard GPS system.

The contribution of the University Paris Diderot for this specific measurement is acknowledged.

We also thank and acknowledge the Geodetic Survey Division (GSD) of Natural Resources Canada (NRCan), for providing the PPP software and the Time Section of the Royal Observatory of Belgium (ROB) for providing the P3 software. Special thanks to F. Lahaye (NRCan) and P. Defraigne (ROB) for the kind support and helpful advice provided on the usage of PPP and P3 software.

We thank also LEICA Geosystems Italia for supplying the geodetic instruments used in surveying the LNGS network. 

Finally, we gratefully thank E. Gschwendtner, N. Ashby and L. Evans for very valuable discussions and support.

\end{document}